# INTERSTELLAR MEDIUM PARAMETERS IN FRONT OF THE EXTERNAL BOW SHOCK

P.A. Sedykh

Irkutsk National Research Technical University.
83, Lermontov str., 664074, Irkutsk, Russian Federation
e-mail: pvlsd@mail.ru

## Abstract

We know that it is the front of the Earth's bow shock where the solar wind kinetic energy flux is transformed into the other kinds the most intensively. In our previous studies, we obtained important relationships that enable calculating the key parameters at transition through the Earth's bow shock front.

One of the most important sources of information on physical processes at the heliosphere boundary are the Voyager 1 and 2 spacecrafts. Since both the solar wind and interstellar medium are supersonic streams, two shocks are formed when flowing around the heliopause. The internal shock, in which the solar wind decelerates to subsonic velocity, is called the heliospheric shock. In the external bow shock, the interstellar gas supersonic flux is decelerated.

The aim of this paper is to generalize the previously obtained equations to the processes in the external bow shock region. If Voyager-1 was equipped with a greater set of measuring instruments, we could have already provided estimations of the interstellar medium key parameters, and described in physical terms what this medium is, using relationships and equations from our studies.

**Key words: external bow shock; interstellar medium; set of measuring instruments; computing of parameters.**

# Introduction

The bow shock (BS) front is the main converter of the solar wind kinetic energy into the electromagnetic energy (Ponomarev, Sedykh, Urbanovich, 2006; Sedykh and Ponomarev, 2012), At transition through the front, the intensity of the tangential component of the solar wind magnetic field and density grow several times. Therefore, among other things, the BS front is a current layer. Since the magnetic field plasma passes through the front, there arises an electrical field in the front coordinate system (Sedykh, 2014). Thus, the BS front is the source of electrical power (Sedykh, 2014). This electrical power is sheared by two sinks (consumers) – the transition layer (TL or magnetosheath) and the magnetosphere. We know the relationships between the magnetic field values ahead and behind the BS. They allow calculating the current on the BS front. According to (Ponomarev, Sedykh & Urbanovich, 2006; Sedykh & Ponomarev, 2012), this electric current value depends on the normal component's jump, and the electric current direction depends on the direction of magnetic field ahead the BS. This current is considerably divergent, i.e. it must close somewhere behind the BS front. This closure may occur in the magnetosphere or in the transition layer. Let's briefly address the origin of the external electric current. The space between BS and magnetopause filled with flux of the solar wind modified plasma. Also, this region converts the solar wind kinetic energy into thermal and electrical energy. Formally, electric current, moving under the BS arc, can be obtained if we take a math operator (**curl**) from the transition layer magnetic field. This field, in turn, is derived from the IMF by converting its components at transition through the oblique bow shock front in accordance with the conservation laws. The physical aspect of the case can be described as follows. The BS front reflects some solar wind protons, forming a current, which is parallel to the front. This electric current can be interpreted as a total drift current of reflected protons. If we take divergence from this electric current, we can obtain an expression for the current density component, which is normal to the bow shock front. There is a potential difference between the BS front and the magnetosphere. This difference is unambiguously (since the TL magnetic field is determined by the solar wind magnetic field) related to the velocity of plasma in the TL. The magnetopause potential is functionally linked to the solar wind parameters (Sedykh, 2015). We followed (Whang, 1987) in the approach to the Earth's bow shock description. All obtained equations in ((Ponomarev, Sedykh, Urbanovich, 2006; Sedykh, 2015) can be applied to calculate key parameters and to model the processes in the bow shock region.

The interstellar medium and solar wind interacting; a complex structure called the heliosphere shock layer is formed. The heliosphere shock layer is multicomponent. In this area, essential roles belong to protons and electrons of the solar wind and interstellar medium, trapped ions, interstellar hydrogen atoms, heliosphere and interstellar magnetic fields, and galactic and anomalous cosmic rays components. In the concept accepted at the moment, the shock layer structure is described by two shocks and a contact surface (Baranov, Krasnobayev, Kulikovsky, 1970). We shall note that according to some papers, in the presence of interstellar magnetic field strong enough, there might be no external bow shock, since the Mach-Alfven number in the interstellar medium is less than 1. In case of partially-ionized interstellar medium, neutral atoms interact with the disturbed plasma component by recharging. The recharge process leads to formation of a so-called "hydrogen wall" – region in front of the heliopause proximal to the interstellar medium with increased concentration of interstellar atoms. The hydrogen wall is composed of secondary atoms formed resulting from recharging of primary interstellar atoms on decelerated interstellar protons.

One of the most important sources of information on physical processes at the heliosphere boundary are the Voyager 1 and 2 spacecrafts. As early as in 2011-2013, new unique data about heliosphere boundary from the Voyager-1, 2 spacecrafts became available. Starting from spring 2011, the Voyager-1 spacecraft has been recording a practically zero speed of the solar wind and during some periods its radial velocity was negative (Krimigis et al., 2011). In August 2012, Voyager-1 revealed a dramatic reduction in fluxes of anomalous cosmic rays component, and a simultaneous increase of cosmic rays galactic component fluxes (Stone et al., 2013) and increased value of magnetic field induction from ~ 0.2 nT to ~ 0.4 - 0.5 nT. However, the magnetic field vector direction has not changed, coinciding with the direction of heliospheric magnetic field. In April-May 2013, the PWS instrument has detected 2.6 kHz radiation. Analysis of this radiation enabled scientists to get a rough estimate of electron density at the generation point of this radiation, which is ~ 0.08 cm$^{-3}$, this exceeds considerably the solar wind density. The analysis of the gradient the frequency varies with, depending on distance, led to the conclusion that Voyager-1 crossed the heliopause on August 25, 2012 (Gurnett et al., 2013).

According to (Zieger et al., 2013), estimates of parameters of plasma in the local interstellar medium indicate that the speed of the interstellar wind is most likely less than both the fast magnetosonic speed and the Alfven speed (but greater than the slow magnetosonic speed). In the study (Zieger et al., 2013) the authors found that the bow shock would take a different form, what is known as a "slow bow shock". The authors say that the Voyager 1 probe is expected to pass through the bow shock well after it leaves the heliosphere, although by then, its battery will be long dead (Zieger et al., 2013*).*

In the paper (Gloeckler and Fisk, 2014) is noted that the Voyager 1 spacecraft is currently in the vicinity of the heliopause, which separates the heliosphere from the local interstellar medium. There has been a precipitous decrease in particles accelerated in the heliosphere and a substantial increase in galactic cosmic rays. The authors devised a test to determine whether or not Voyager 1 has left the heliosphere. Previously, researchers found that Voyager 1 has encountered increased density of electron plasma, which they interpret as evidence that Voyager 1 is in interstellar space. According to (Gloeckler and Fisk, 2014) if Voyager 1 encounters another current sheet, it will provide strong observational evidence that the heliopause has not yet been crossed. Using data about Voyager 1's current location and speed and the dynamics of the Sun's current sheets, the authors determined that the spacecraft should encounter another current sheet during 2015 and certainly by no later than the end of 2016 (Gloeckler and Fisk, 2014).

The main purpose of the paper is to study physical processes at the heliosphere boundary (the region of the solar wind interaction with the interstellar medium) by means of theoretical analysis of some experimental data. With the Voyagers and IBEX returning many new puzzles, this paper could be interesting because it might address in a complementary way questions that are hotly debated in the heliophysics community.

**Some application of the derived equations for the external bow shock**

Baranov et al. (1970) proposed a structure with two shocks, which is now the basis of a concept of the heliosphere shock layer (see Fig. 1). Heliopause, which is a tangential discontinuity surface, separates the interstellar medium charged component from the plasma of the solar wind. Because both the solar wind and interstellar medium are supersonic streams, two shocks are formed when flowing around the heliopause: heliospheric shock and external bow shock.

It is known that important relations and equations to calculate **thermodynamical** parameters at transition through the shock front are named in recognition of the work carried out by Scottish physicist William John Macquorn Rankine and French engineer Pierre Henri Hugoniot. The Rankine-Hugoniot relations apply to a one-dimensional planar shock. Attempting to apply them to the geometry of the bow shock is not straight forward since the shape of the shock is not a priori known. In fact, the Rankine- Hugoniot relations must be applied in conjunction with calculating the shock shape, the downstream flow, the heliopause shape, the inner-heliosheath configuration, and the termination shock location. The Rankine-Hugoniot relations are based on energy conservation laws, which have been generalized for different shocks. The interstellar medium is most likely a multi-ion plasma containing both protons and He ions, and the heliosheath is know to contain both thermal solar wind ions, hot pickup ions as well as anomalous and galactic cosmic ray particles, which are not in thermodynamic equilibrium. The local interstellar medium also contains neutral atoms, which interact with the solar wind plasma through charge exchange, electron impact ionization or photoionization. Due to all these complications, one could apply the classical Rankine-Hugoniot relations only with combinations of equations from our studies; and one should take into account the results of recent computer simulations.

Let's also address other key parameters. We can choose the system of coordinates as a base beginning at the center of the Sun (Fig.1). I shall use local coordinate system (l, k, n). I will follow our previously published papers (Ponomarev, Sedykh, Urbanovich, 2006; Sedykh 2011; Sedykh, 2014) in the approach to the external bow shock description.

I assume a spherically symmetric heliospheric bow shock like for the Earth, while there is hypothesis, since the Voyager passages of the termination shock and through IBEX observations, that the heliosphere might be strongly distorted. I need to make such simplification to solve the problem analytically.

The correspondences of the parameters in front of and behind (in TL) the external bow shock will be the following. The plasma density is defined as:

$$\rho_2 = \rho_1 \cdot \delta; \quad \delta = \frac{\gamma+1}{\gamma-1};$$

One can derive the gas pressure:

$$p_{2g} = 2\rho_1 V_1^2 \frac{\sin^2(\alpha)}{\gamma+1}$$

The tangential velocity component will be:
$$V_{2l} = V_{1l} = V_1 \cos(\alpha)$$
The normal velocity component will be:
$$V_{2k} = \frac{V_{1k}}{\delta} = V_1 \frac{\sin(\alpha)}{\delta}$$
The vertical component of the magnetic field will be:
$$B_{2n} = B_{1n} \cdot \delta$$
The tangential component of the magnetic field is defined as:
$$B_{2l} = B_{1l} \cdot \delta = B_{1eq} \cdot \delta \cos(\alpha + \beta)$$
The normal component of the magnetic field is defined as:
$$B_{2k} = B_{1k} = B_{1eq} \sin(\alpha + \beta)$$

In these equations, numbers 1, 2 correspond to the interstellar medium and the transition layer, respectively; γ – is the adiabatic exponent; α – is the angle between the tangent to the external bow shock front and the X-axis; β – is the angle between the direction of the interstellar wind velocity and the projection of the magnetic field onto the equatorial plane $B_{1eq}$ (see Figure 1).

Besides, knowing numerical values of parameters in the equations, we can set various values of γ. Thus, one can obtain the additional information on properties of medium.

We should note that owing to final curvature of the external bow shock front surface there appears a force (additional) of magnetic tension: $F_n = (B\nabla)B_n/4\pi$. This force must be put in equilibrium by additional Ampere force: $j^*_\tau B_l/c$; here $j^*_\tau$ is density of additional current: $j^*_\tau = c \cdot (\partial B_n/\partial l)/4\pi$. We can find the density of the basic current $j_\tau$: $j_\tau = c (\delta - 1)B_{0l}/d$, where d is front thickness. Since d <<$X_h$, $X_h$ - is the distance from the origin of system of coordinates to the shock front, then additional electric current is much less than basic one. The effect of curvature should not be taken into account, at least until δ notably differs from

1. We can obtain the equation for the gradient of plasma pressure and the inertial force behind the external bow shock front (i.e. in the transitional layer):

$$\frac{dP_{2g}}{dl} = \frac{2\rho_1 V_1^2}{\gamma+1}\frac{d}{dl}\sin^2\alpha \; ; \quad \frac{\rho_2}{2}\frac{dV_2^2}{dl} = -\frac{\rho_1 V_1^2(\delta^2-1)}{\delta}\frac{d}{dl}\sin^2\alpha \; ; \quad (1)$$

where $\delta = (\gamma+1)/(\gamma-1)$.

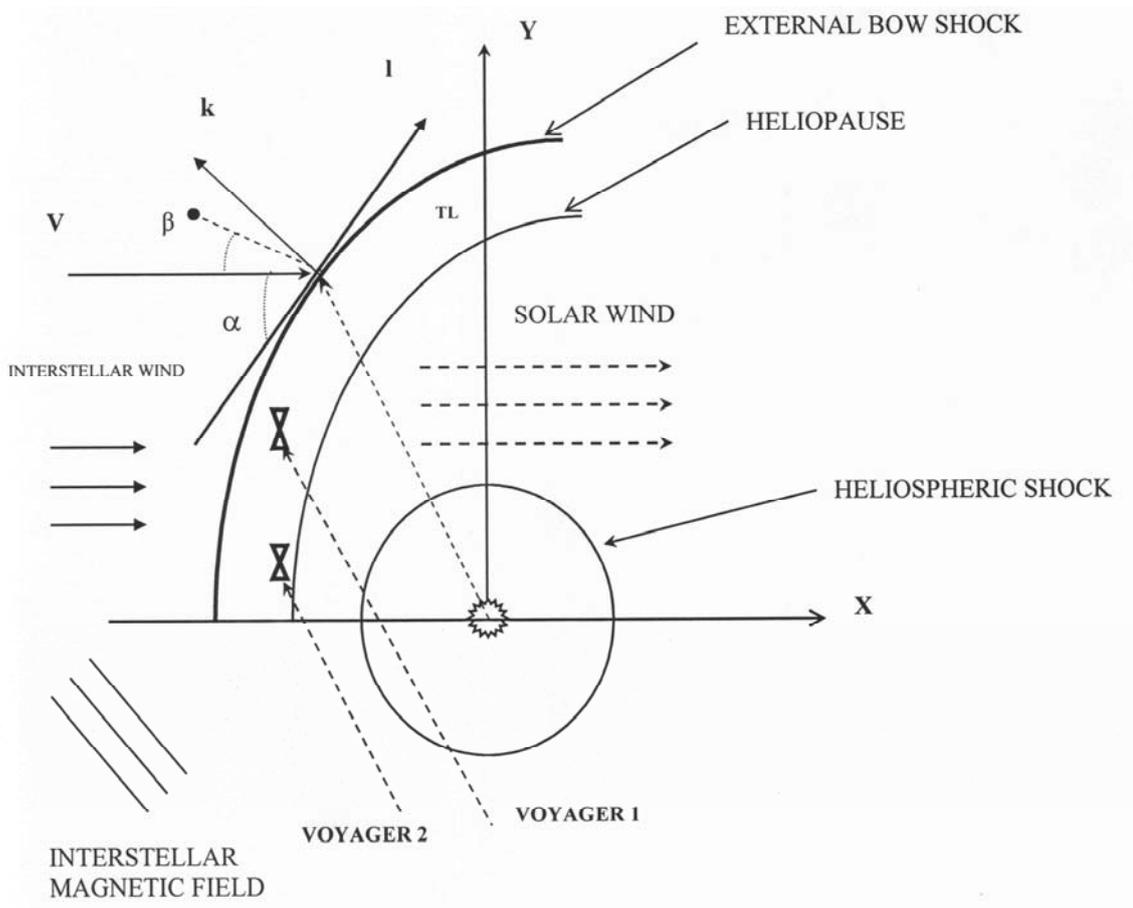

Figure 1. A sketch of the external bow shock, of the Transition Layer (TL), heliopause. The location of the orthogonal coordinate system (X,Y) with the origin in the Sun's center and the location of the local coordinate system (l, k, n); axis l is directed along the tangent to the external bow shock front, axis k is directed perpendicularly to the tangent (to the external bow shock front) and axis n is directed perpendicularly to the sketch plane (i.e. the n- axis supplements this system to the right one).

$\alpha$ - is an angle between the tangent to the external bow shock front and axis X; $\beta$ - is an angle between the interstellar wind velocity and projection of interstellar magnetic field onto the equatorial plane $B_{1eq}$.

Thus, substituting these expressions into the equation for current density, we can obtain:

$$j = \frac{c}{B_2^2}\left[B \times \left(\nabla p_g + \rho \frac{\nabla V^2}{2}\right)\right]; \quad j_{2k} = c\left(\frac{dP_{2g}}{dl} + \rho \frac{dV_{2l}^2}{dl}\right)\frac{B_{2n}}{B_2^2}; \text{ or}$$

$$j_{2k} = -\frac{2B_{1n}}{(\gamma^2 - 1)B_2^2} c\delta \rho_1 V_1^2 \frac{d}{dl}\sin^2\alpha, \qquad (2)$$

where $j_{2k}$ – is an electric current, which flows across the transition layer;

$$B_2^2 = B_{1n}^2 \delta^2 + B_{1eq}^2\left((\delta^2 + 1) - 2(\delta^2 - 1)\sin\alpha \cdot \cos\alpha\right); \qquad (3)$$

The surface density of the current, flowing in the transitive layer along it, will be the integral from $j_l$ across the layer from heliopause up to the external bow shock: $J_l = cB_{1n}\delta \frac{P_g(bs) - P_g(h)}{B_2^2}$; in this expression $P_g(bs)$ and $P_g(h)$ are gas pressure under the external bow shock and on the heliopause, respectively.

One can obtain the important relationship for the physics of the shock layer, in which the hydrodynamic and electrodynamic quantities are in the left and right sides of this equation, respectively:

$$V_e \nabla P_g^e + V_i \nabla P_g^i = Ej - \rho(V_i - V_e)\frac{dV_i}{dt},$$

where e – index for electrons; i – index for ions.

A normal component of velocity of medium on border is equal to zero; hence, the flow of the number of particles, transferable by the electric current, will be $N_{TL} = j_{2k}/2e$; e – is the charge of electron.

We can also assume, that the front form of the external bow shock is given, as well as the form of heliopause. Heliopause may be well approximated by biaxial hyperboloid. One can assume that both heliopause and external bow shock are paraboloids of rotation and differ only in various distances to the nose point. Such step is connected with the fact that the form of heliopause and especially forms of the external bow shock front differ little from paraboloids of rotation, at the same time all analytical expressions drastically become simpler (see corresponding equations in (Madelung, 1957)). Parabola is some compromise between ellipsoid (closed model) and hyperboloid (open model). Further, we can apply important relationships from our papers that enable calculating the key parameters at transition through the external bow shock front.

The system 'external bow shock-TL-heliopause-heliospheric shock' is unique plasma laboratory. Thus, if we knew such parameters e.g. as plasma density, plasma pressure, gas pressure gradient, components of magnetic field, electric field, fluxes of particles in the transition layer, we would be able to determine key parameters in front of the external bow shock.

## Conclusion

To obtain the analytically solution of the problem, we have been compelled to make some simplifications (by analyzing a maximum possible simple model that at the same time retains the most important traits of reality). In most cases, we followed (Baranov, Krasnobayev, Kulikovsky, 1970) in the approach to the heliosphere shock layer description. In this paper, we use the results of previous studies, in which the following was obtained: the expressions for the magnetic field components at transition through the Earth's bow shock front; the equations for plasma density; the expressions for electrical current generated in the bow shock front and closed through the Earth's magnetosphere, and for the magnetopause potential as a function of the solar wind parameters – plasma density, velocity and intensity of the interplanetary magnetic field (IMF); and also, the expressions for the transition layer basic parameters as functions of the IMF components etc.

We can use the obtained important expressions and relationships to determine the interstellar medium parameters. It is known that the Voyager-1 spacecraft was equipped with the following scientific instruments: UV spectrometer, interference IR spectrometer, photopolarimeter, low-energy charged particle detector, instrument to determine radio waves of planets, instrument to determine waves in plasma, magnetometer to measure weak magnetic fields, magnetometer to measure strong magnetic fields, cosmic ray detector, and plasma detector.

If Voyager-1 was equipped with a broader range of measuring instruments, we could have already provided estimations of the interstellar medium key parameters, using the above mentioned relationships, equations. In other words, if we knew the parameters in the transition layer, we would be able to calculate them ahead the external bow shock front. Thus, we could have now made preliminary conclusions on the behind-the-heliosphere medium, i.e. in the interstellar space. It should be noted that even if the Voyager-1,2 spacecrafts were initially equipped with a great set of measuring instruments, they would need to be protected against cosmic radiation. Otherwise, cosmic ionizing radiation would destroy almost all the electronics inside the apparatus on their way out of the solar system.

I have provided the specific equations. We have all the essential equations to calculate the parameters; we have a developed mathematical apparatus to convert the relevant physical values at transition through the bow shock front, a set of computer programs with the user interface; two spacecrafts are in the appropriate region, but they do not have the necessary measuring instruments. This essential set of measuring instruments that would enable us to obtain a series of numerical estimates of key parameters can be presented as follows:

**1. Magnetometer. FluxGate Magnetometer (Magnetic Field Vector, spin resolution; Magnetic Field Magnitude).**

**2. Ion and electron sensor combination that operates in the bulk plasma energy regime/Instrument to measure plasma density and composition (Electron, Proton, and Alpha-particle Monitor).**

**3. Electron Diagonalized Temperature. Electron Symmetry Vector. Ion Diagonalized Temperature. Ion Symmetry Vector.**

**4. Electrostatic Analyzer.**

**5. Electric Field Variation, based on spin plane component.**

We should note that to date, this set of measuring instruments is not a hard-to-obtain one, it is readily available on research satellites. There is bound to be a huge time-lag before Voyager spacecrafts can travel through the interstellar space medium, and we could already be able to learn this medium properties, because the parameters behind the shock front are well-determined by the known laws and relationships. Parameters of the medium behind the front of the external bow shock contain much information about physical parameters ahead of the external bow shock front (about interstellar medium). Any follow-up mission to the heliospheric boundary that would be worth the effort and that a space agency would be willing to fund, which could carry this instrumentation, almost for sure would be an interstellar probe that continues into the interstellar medium proper. This would make most of the motivation for similar studies.

The proposed study would significantly reduce the current uncertainties concerned with the structure of the heliosphere shock layer behind the external bow shock, and with measuring the parameters of the local interstellar medium that surrounds the solar system.

**The research of this paper was funded by the programs of Irkutsk National Research Technical University.**
**This is a theoretical investigation; data supporting this [SYSTEMATIC REVIEW or META-ANALYSIS] are from previously reported studies, which have been cited.**